\documentclass[twocolumn,showpacs,floatfix,superscriptaddress,amsmath,amssymb,prl]{revtex4}
\usepackage{txfonts}
\usepackage{amssymb}
\usepackage{graphicx}
\usepackage{hyperref}
\usepackage{ulem}
 \usepackage{overpic}
 \usepackage{psfrag}
\newcommand{\be}{\begin{equation}}
\newcommand{\ee}{\end{equation}}

\begin{document}

\title{Singularities in fidelity surfaces for quantum phase transitions: a geometric perspective}

\author{Huan-Qiang Zhou}
\affiliation{Centre for Modern Physics and Department of Physics,
Chongqing University, Chongqing 400044, The People's Republic of
China}
\author{Jian-Hui Zhao}
\affiliation{Centre for Modern Physics and Department of Physics,
Chongqing University, Chongqing 400044, The People's Republic of
China}
\author{Hong-Lei Wang}
\affiliation{Centre for Modern Physics and Department of Physics,
Chongqing University, Chongqing 400044, The People's Republic of
China}
\author{Bo Li}
\affiliation{Centre for Modern Physics
and Department of Physics, Chongqing University, Chongqing 400044,
The People's Republic of China}
\begin{abstract}
The fidelity per site between two ground states of a quantum lattice
system corresponding to different values of the control parameter
defines a surface embedded in a Euclidean space. The Gaussian
curvature naturally quantifies quantum fluctuations that destroy
orders at transition points. It turns out that quantum fluctuations
wildly distort the fidelity surface near the transition points, at
which the Gaussian curvature is singular in the thermodynamic limit.
As a concrete example, the one-dimensional quantum Ising model in a
transverse field is analyzed.  We also perform a finite size scaling
analysis for the transverse Ising model of finite sizes. The scaling
behavior for the Gaussian curvature is numerically checked and the
correlation length critical exponent is extracted, which is
consistent with the conformal invariance at the critical point.

\end{abstract}

\date{\today}
\maketitle

Quantum phase transitions (QPTs) have been a research topic subject
to intense study,  since their significant role was realized in
accounting for high-$T_c$ superconductors, fractional quantum Hall
liquids, and quantum magnets~\cite{sachdev,wen}. Recently,
significant advances have been made in attempt to clarify the
connection between quantum many-body physics and quantum information
science. This provides a new perspective to investigate QPTs from
\textit{entanglement}~\cite{osborne,vidal,levin,entanglement1,entanglement2}
and \textit{fidelity}~\cite{zanardi,zjp,zhou,zov,more}, basic
notions of quantum information science~\cite{nielsen} and turns out
to be very insightful in our understanding of QPTs in a variety of
quantum lattice systems in condensed matter.

Conventionally, orders and fluctuations provide a proper language to
study QPTs, with order parameters being the key to quantify quantum
fluctuations. Instead, the fidelity approach is based on state
distinguishability arising from the orthogonality of different
ground states in the thermodynamic limit. In fact, the ground state
fidelity for a quantum system may be mapped onto the partition
function of the equivalent classical statistical lattice model with
the same geometry~\cite{zov}. Thus, the fidelity per site is
well-defined in the thermodynamic limit, and its singularities
unveil transition points, at which the system under consideration
undergoes QPTs. Therefore, a practical means is now available to map
out the ground state phase diagram for a quantum lattice system
without prior knowledge of order parameters. An intriguing question
is how to characterize singularities in the fidelity per site.
Indeed, a proper answer to this question will shed new light on our
understanding of QPTs.

In this paper, we present an \textit{intrinsic} characterization of
singularities in the fidelity per site in terms of Riemannian
geometry. For this purpose, we first \textit{define} a fidelity
surface as a surface embedded in a Euclidean space,  which in turn
is determined by the average fidelity per lattice site between two
ground states of a quantum lattice system as a function of the
control parameters. This makes the whole machinery developed in
differential geometry of curves and surfaces available to study
QPTs. As it is well known, the Gaussian curvature, or equivalently,
the Ricci scalar curvature for the surfaces embedded in Euclidean
spaces, is a fundamental concept used to measure how curved a
surface is. Therefore, the Gaussian curvature is expected to
naturally quantifies quantum fluctuations that destroy orders at
transition points.  We discuss the global behaviors of the Gaussian
curvature. It turns out that quantum fluctuations wildly distort the
fidelity surfaces near the transition points. Generically,
precursors of QPTs occur in the Gaussian curvature for finite-size
systems. In the thermodynamic limit, the Gaussian curvature becomes
singular at transition points. The one-dimensional quantum Ising
model in a transverse field is exploited to explicitly illustrate
the theory. We also perform a finite size scaling analysis for the
Gaussian curvature with different lattice sizes to extract the
correlation length critical exponent.

{\it Fidelity surfaces.} For a quantum lattice system described by a
Hamiltonian $H(\lambda)$, with $\lambda$ a control parameter. Here
we restrict ourselves to discuss the simplest case with one single
control parameter, although the extension to multiple control
parameters is straightforward. For two ground states
$|\psi(\lambda_1)\rangle$ and $|\psi(\lambda_2)\rangle$
corresponding to different values of the control parameter
$\lambda$, the fidelity is defined as $F(\lambda_1, \lambda_2)\equiv
|\langle\psi(\lambda_2)|\psi(\lambda_1)\rangle|$. For a large but
finite $L$, the fidelity $F$ asymptotically scales as $F(\lambda_1,
\lambda_2) \sim d^L(\lambda_1, \lambda_2)$, where the scaling
parameter $d(\lambda_1, \lambda_2)$ characterizes how fast the
fidelity changes when the thermodynamic limit is
approached~\cite{zhou}. Physically, it is the fidelity per site.
Here note that the contribution from each site to $F(\lambda_1,
\lambda_2)$ is multiplicative. Following~\cite{zov}, the ground
state fidelity for a quantum system is nothing but the partition
function of the equivalent classical statistical lattice model with
the same geometry, if one utilizes the tensor network
representations of ground state many-body wave functions. Therefore,
$d(\lambda_1,\lambda_2)$ may be interpreted as the partition
function per site~\cite{baxter}, which is well-defined in the
thermodynamic limit:
\begin{equation} \ln d(\lambda_1,\lambda_2) = \lim_{L
\rightarrow \infty} \ln F(\lambda_1,\lambda_2) /L.
\end{equation}
The fidelity per site $d(\lambda_1,\lambda_2)$ satisfies the
properties: (1) symmetry under interchange $\lambda_1
\longleftrightarrow \lambda_2$; (2) $d(\lambda_1,\lambda_1)=1$; and
(3) $0\leq d(\lambda_1,\lambda_2) \leq 1$.

For simplicity, let us assume that the system undergoes a QPT at
$\lambda_c$. If $|\psi(\lambda_1)\rangle$ and
$|\psi(\lambda_2)\rangle$ are in the same phase, then they flow to
the same stable fixed point in the sense of renormalization group
theory, and so their difference arises from quantum fluctuations
depending on the details of the system. On the other hand, if
$|\psi(\lambda_1)\rangle$ and $|\psi(\lambda_2)\rangle$ are in
different phases, then they flow to two different stable fixed
points. Therefore, they possess different orders, although quantum
fluctuations originate from the same unstable fixed point
$\lambda_c$~\cite{relevant}. Imagine that if there were no quantum
fluctuations, then $d(\lambda_1,\lambda_2)$ would be simply 1 when
$|\psi(\lambda_1)\rangle$ and $|\psi(\lambda_2)\rangle$ are in the
same phase; otherwise, when $|\psi(\lambda_1)\rangle$ and
$|\psi(\lambda_2)\rangle$ are in different phases,
$d(\lambda_1,\lambda_2)$ would take the minimum value corresponding
to the two stable fixed points. For continuous QPTs, quantum
fluctuations are strong enough such that no orders survive at the
transition point, so $d(\lambda_1,\lambda_2)$ is continuous, but
displays singularities, whereas for the first order QPTs,
$d(\lambda_1,\lambda_2)$ remains to be discontinuous at the
transition point.  An interesting observation is to regard the
fidelity per site, $d(\lambda_1,\lambda_2)$, as a two-dimensional
surface embedded in the three-dimensional Euclidean space, with a
Riemannian metric induced from the Euclidean metric. Our aim is to
give an \textit{intrinsic} characterization of singularities in such
a fidelity surface in terms of Riemannian geometry.

{\it Differential geometry of the two-dimensional surfaces embedded
in the three-dimensional Euclidean space.} Let us briefly recall the
fundamentals of differential geometry of surfaces embedded in
Euclidean spaces~\cite{novikov}. For a two-dimensional surface
embedded in a three-dimensional Euclidean space:
$z=f(\lambda_1,\lambda_2)$, the first fundamental form on the
surface is
\begin{equation}
dl^{2}=g_{ij}d\lambda^{i}d\lambda^{j}=E(du)^{2}+2F(du
dv)+G(dv)^{2}\label{frist},\end{equation} where $g_{ij}$ is the
Riemannian metric on the surface: $g_{11}=1+f_{\lambda_1}^{2}$,
$g_{12}=g_{21}=f_{\lambda_1}f_{\lambda_2}$, and
$g_{22}=1+f_{\lambda_2}^{2}$. Here the subscripts $\lambda_1$ and
$\lambda_2$ denote partial differentiations with respect to
$\lambda_1$ and $\lambda_2$, respectively. In terms of the
co-ordinates $u=\lambda_1$ and $v=\lambda_2$, we have $E=g_{11}$,
$F=g_{12}=g_{21}$ and $G=g_{22}$. Suppose the surface is given in
parametric form: $r=r(u, v)$. Then, the vector product $r_{u} \times
r_{v}$ is a non-zero vector perpendicular to the surface at each
non-singular point; define $m$ to be a unit vector in the normal
direction, then one has $r_{u} \times r_{v} =|r_{u} \times r_{v}|m$.
For a curve $r=r(u(l),v(l))$ on the surface, the projection of the
second order derivative $\ddot{r}$ of $r$ with respect to the arc
length $l$ on the normal to the surface leads to the second
fundamental form as follows
\begin{equation}
\langle \ddot{r},m \rangle
(dl)^2=b_{ij}d\lambda^{i}d\lambda^{j}=X(du)^{2}+2Y du dv+Z (dv)^{2},
\label{second}
\end{equation}
if a surface is given in the form $z=f(\lambda_1,\lambda_2)$ with
$\lambda_1=u$, $\lambda_2=v$, and $r(u,v)=(u,v,f(u,v))$. Therefore,
we have $X=b_{11}=f_{\lambda_1\lambda_1}/ \sqrt
{1+f_{\lambda_1}^{2}+f_{\lambda_2}^{2}}$,
$Y=b_{12}=b_{21}=f_{\lambda_1\lambda_2}/ \sqrt
{1+f_{\lambda_1}^{2}+f_{\lambda_2}^{2}}$ and
$Z=b_{22}=f_{\lambda_2\lambda_2}/ \sqrt
{1+f_{\lambda_1}^{2}+f_{\lambda_2}^{2}}$.

The eigenvalues of the pair of quadratic forms (\ref{frist}) and
(\ref{second}) are the {\it principal curvatures} of the surface at
the point under investigation. The product of the principal
curvatures is the {\it Gaussian curvature} $K$ of the surface at the
point, and their sum the {\it mean curvature}. The principal
curvatures $k_{1}$ and $k_{2}$ are the solutions of equation:
\begin{equation}
{\rm det}(Q-k G)=0 \label{det}, \end{equation} where $Q=(b_{ij})$ is
the matrix of the second fundamental form, and $G=(g_{ij})$. Since
the first fundamental form is positive definite, its matrix $G$ is
non-singular. Hence ${\rm det}(Q-k G)={\rm det} G \; {\rm
det}(G^{-1}Q - k \cdot I)$, we deduce that the Gaussian curvature
$K=k_{1} k_{2}={\rm det}(G^{-1}Q)={\rm det}Q / {\rm det}G$ and the
mean curvature $M=k_{1}+k_{2}={\rm tr}(G^{-1} Q)$. Therefore, the
Gaussian curvature $K$ and the mean curvature $M$ take the form:
\begin{equation} K=\frac {f_{\lambda_1\lambda_1} f_{\lambda_2\lambda_2}-
f_ {\lambda_1 \lambda_2}^{2}}{\left(1 + f_{\lambda_1}^{2} +
f_{\lambda_2}^{2}\right)^{2}}, \label{gaussian curvature}
\end{equation}
and
\begin{equation} M=\frac {(1+f_{\lambda_2}^{2})f_{\lambda_1\lambda_1}+ (1+f_{\lambda_1}^{2})f_{\lambda_2\lambda_2}-
2f_{\lambda_1}f_{\lambda_2}f_ {\lambda_1 \lambda_2}^{2}}{\left(1 +
f_{\lambda_1}^{2} + f_{\lambda_2}^{2}\right)^{\frac{3}{2}}},
\label{mean curvature}
\end{equation}
respectively. We notice that the sign of the {\it Gaussian
curvature} $K$ is the same as the sign of the determinant:
$f_{\lambda_1 \lambda_1} f_{\lambda_2\lambda_2}- f_
{\lambda_1\lambda_2}^{2}$, i.e., the {\it Hessian} of $z=
f(\lambda_1,\lambda_2)$

It follows that, in contrast with the mean curvature $M$, the
Gaussian curvature $K$ of a surface may be expressed in terms of the
induced metric on the surface alone, and is therefore an intrinsic
invariant of the surface~\cite{novikov}. In addition, a
two-dimensional surface in a three-dimensional space may also be
regarded as a differentiable manifold endowed with a Riemannian
metric induced from the Euclidean metric. The Ricci scalar curvature
$R$ is twice the Gaussian curvature $K$: $R=2K$.

{\it Global behaviors of the Gaussian curvature $K$ for a fidelity
surface.} Now we consider the (logarithmic function of) fidelity per
site, $\ln d(\lambda_1,\lambda_2)$, as a two-dimensional surface
embedded in the three-dimensional Euclidean space:
$z=f(\lambda_1,\lambda_2) \equiv \ln d(\lambda_1,\lambda_2)$.  The
Gaussian curvature $K(\lambda_1,\lambda_2)$ for such a fidelity
surface may be used to quantify how strong quantum fluctuations are
in given quantum many-body ground states, thus providing an
intrinsic characterization of singularities in the fidelity surface.
Indeed, as justified in Refs.~\cite{zjp,zhou,zov}, the fidelity per
site $d(\lambda_1,\lambda_2)$  is singular when $\lambda_1
(\lambda_2)$ crosses $\lambda_c$ for a fixed $\lambda_2 (\lambda_1)$
in the thermodynamic limit. Therefore the Gaussian curvature
$K(\lambda_1,\lambda_2)$ for the fidelity surface is singular at
$\lambda_1=\lambda_c$ and/or $\lambda_2=\lambda_c$ in the
thermodynamic limit. Generically, we have: $(1)$
$K(\lambda_1,\lambda_2)>0$, there is a neighborhood of the point
throughout which the surface lies on one sides of the tangent plane
at the points; $(2)$ $K(\lambda_1,\lambda_2)<0$, then the surface
intersects the tangent plane at the point arbitrarily close to the
point. If the surface is strictly convex, then we say that the
Gaussian curvature $K(\lambda_1,\lambda_2)$ is positive at every
point of the surface. That is what happens if $\lambda_1$ and
$\lambda_2$ are away from the transition point.  However, if
$\lambda_1$ and $\lambda_2$ are close to the transition point, then
the Gaussian curvature $K(\lambda_1,\lambda_2)$ can be negative.

For finite-size systems, the Gaussian curvature
$K(\lambda_1,\lambda_2)$ remains to be smooth, although the
precursors of QPTs occur as anomalies in the Gaussian curvature
$K(\lambda_1,\lambda_2)$. The anomalies get more pronounced when the
thermodynamic limit is approached. We may take advantage of this
fact to perform finite size scaling to extract the correlation
length critical exponent.
\begin{figure}[ht]
\begin{overpic}[width=72mm,totalheight=42mm]{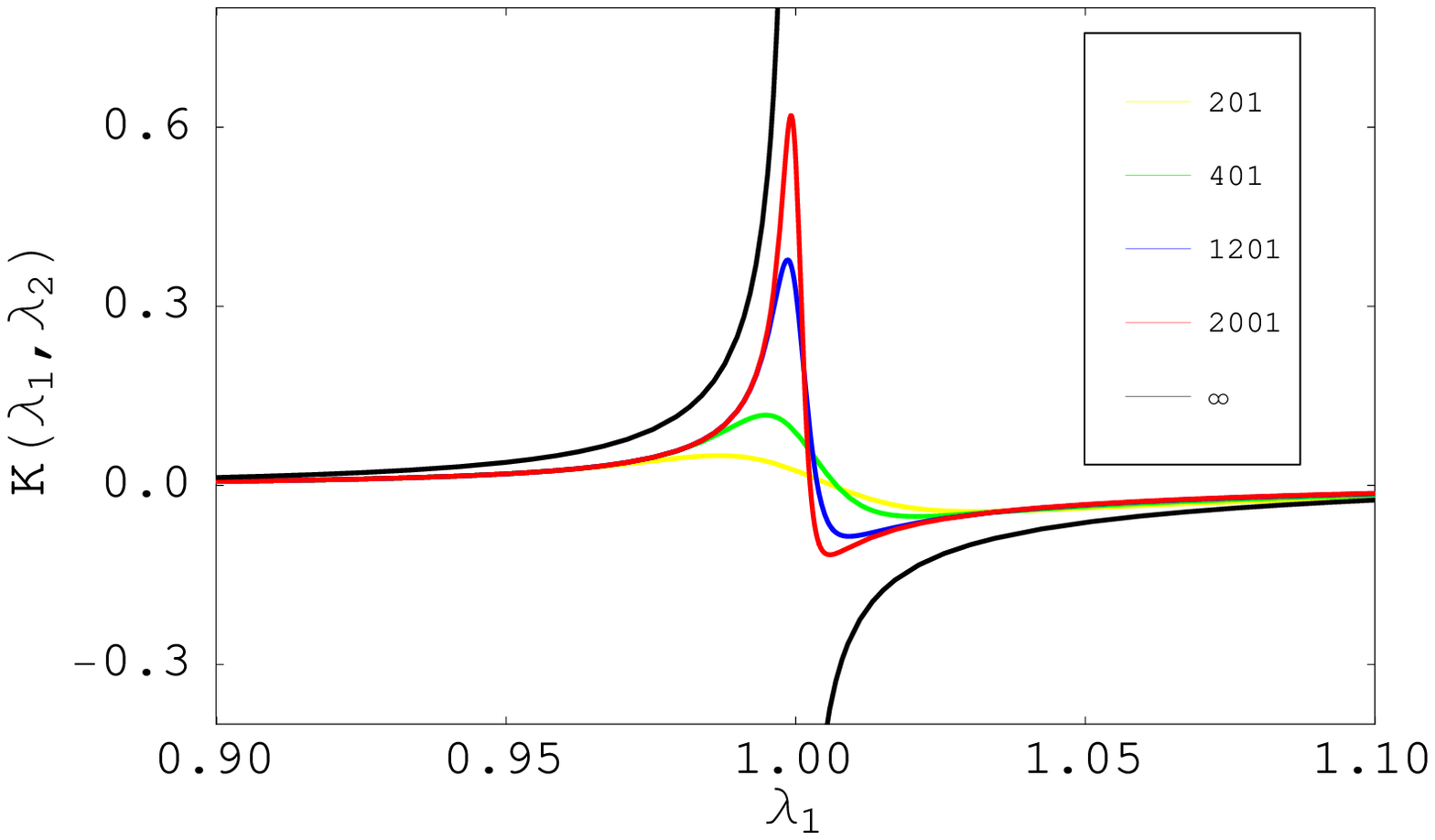}
\put(30,12){$(a)$}
\end{overpic}
\hspace{0in}
\begin{overpic}[width=72mm,totalheight=42mm]{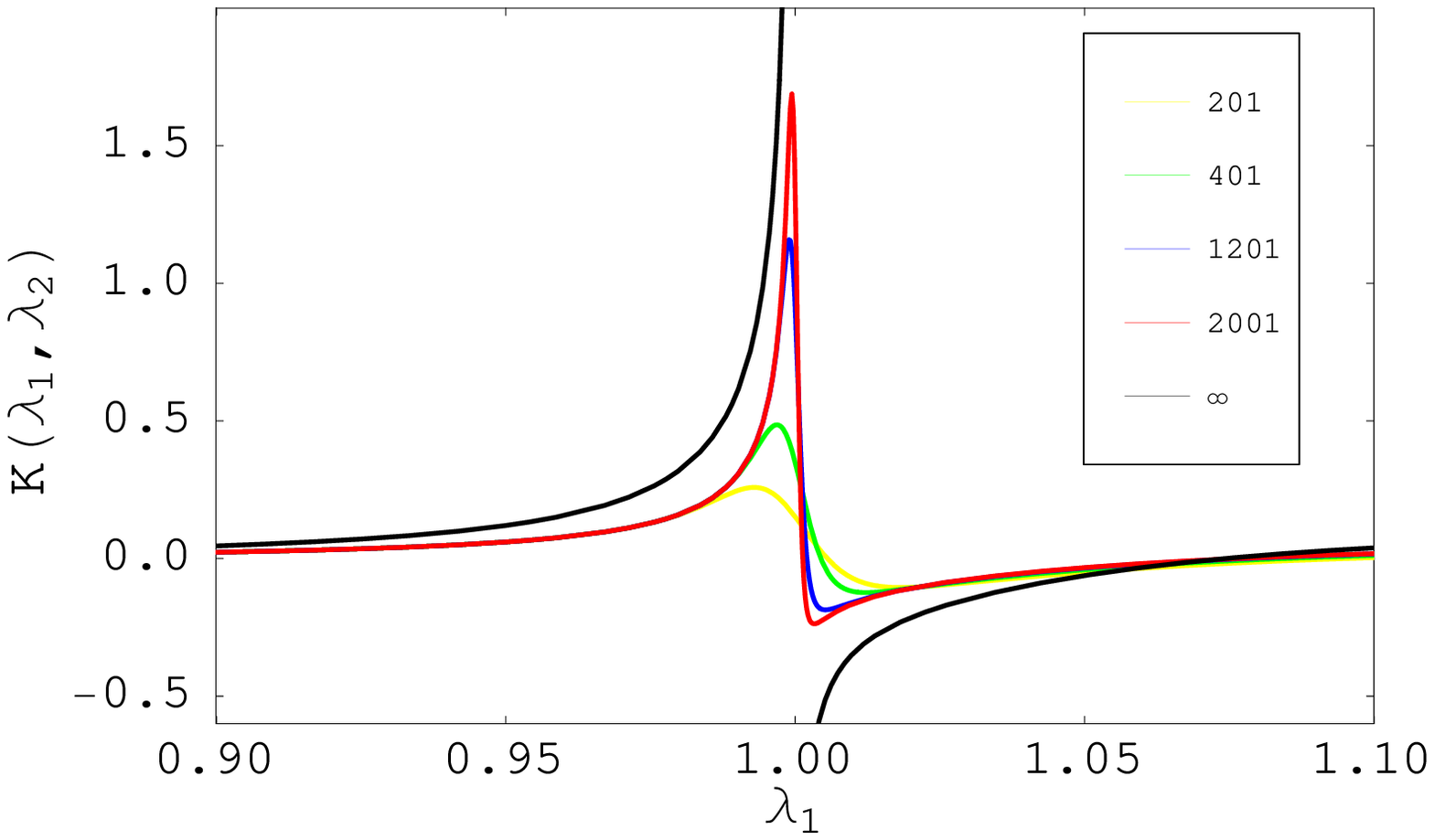}
\put(30,12){$(b)$}
\end{overpic}
  \caption{(color online) The behavior near the critical
  point $\lambda_c=1$ is analyzed for the Gaussian curvature $K(\lambda_1,\lambda_2)$ of the quantum transverse Ising model
  for various lattice sizes.
  The curves shown correspond to different lattice sizes $L=201,401,1201,2001$, and $\infty$.
  The peaks (dips) get more pronounced in the left (right) side with increasing system size.
  The Gaussian curvature
$K(\lambda_1,\lambda_2)$ diverges at the critical point $\lambda_1 =
\lambda_c$ for the infinite-size system ($L= \infty$). Upper panel:
Here  $K(\lambda_1,\lambda_2)$ is regarded as a function of
$\lambda_1$ for $\lambda_2 =0.6$ and $\gamma =1$. Lower panel: Here
$K(\lambda_1,\lambda_2)$ is regarded as a function of $\lambda_1$
for $\lambda_2 =0.6$ and $\gamma =1/2$.} \label{fig1}
\end{figure}

{\it Quantum $XY$ spin 1/2 model.} The quantum $XY$ spin model is
described by the Hamiltonian
\begin{equation}
H= -\sum_{j=-M}^M \left( \frac {1+\gamma}{2} \sigma^x_j
\sigma^x_{j+1} + \frac {1-\gamma}{2} \sigma^y_j \sigma^y_{j+1}  +
\lambda \sigma^z_j \right). \label{HXY}
\end{equation}
Here $\sigma_j^x, \sigma_j^y$ and $\sigma_j^z$ are the Pauli
matrices at the $j$-th lattice site. The parameter $\gamma$ denotes
an anisotropy in the nearest-neighbor spin-spin interaction, whereas
$\lambda$ is an external magnetic field. The Hamiltonian (\ref{HXY})
may be exactly diagonalized~\cite{lieb,pfeuty} for any finite size
$L$ with $L=2M+1$. In the thermodynamic limit $L \rightarrow
\infty$, $\ln d(\lambda_1,\lambda_2)$ takes the form~\cite{zjp}:
\begin{equation}
\ln d(\lambda_1,\lambda_2) = \frac {1}{2\pi} \int ^\pi_0 d\alpha \ln
{\cal F} (\lambda_1,\lambda_2;\alpha),\label{dforising}
\end{equation}
where ${\cal F}(\lambda_1,\lambda_2;\alpha)=\cos [\vartheta
(\lambda_1;\alpha)-\vartheta(\lambda_2;\alpha)]/2,$ with $\cos
\vartheta(\lambda;\alpha)=(\cos \alpha - \lambda)/\sqrt {(\cos
\alpha -\lambda)^2+\gamma^2 \sin^2 \alpha}$~\cite{finitefidelity}.

Now it is straightforward to calculate the Gaussian curvature
$K(\lambda_1,\lambda_2)$ for the fidelity surface of the quantum
$XY$ spin chain. In Fig.~\ref{fig1}, we plot
$K(\lambda_1,\lambda_2=0.6)$ for the fidelity surface of the quantum
$XY$ model ($\gamma=1$ for the upper panel and $\gamma=1/2$ for the
lower panel). One observes that $K(\lambda_1,\lambda_2=0.6)$ is
divergent as a function of $\lambda_1$ at the critical point
$\lambda_c=1$ for the infinite-size system $L=\infty$, indicating
that the fidelity surface is wildly distorted, due to strong quantum
fluctuations near the critical point. This is true for any nonzero
$\gamma$, consistent with the fact that the quantum $XY$ model for
any nonzero $\gamma$ belongs to the same universality class as the
quantum transverse Ising model. That is, there is a critical line
$\gamma\neq 0$ and $\lambda_c=1$; only one (second-order) critical
point $\lambda_c=1$ separates two gapful phases: (spin reversal)
$Z_2$ symmetry-breaking and symmetric phases.

\begin{figure}[ht]
\begin{overpic}[width=42mm,totalheight=42mm]{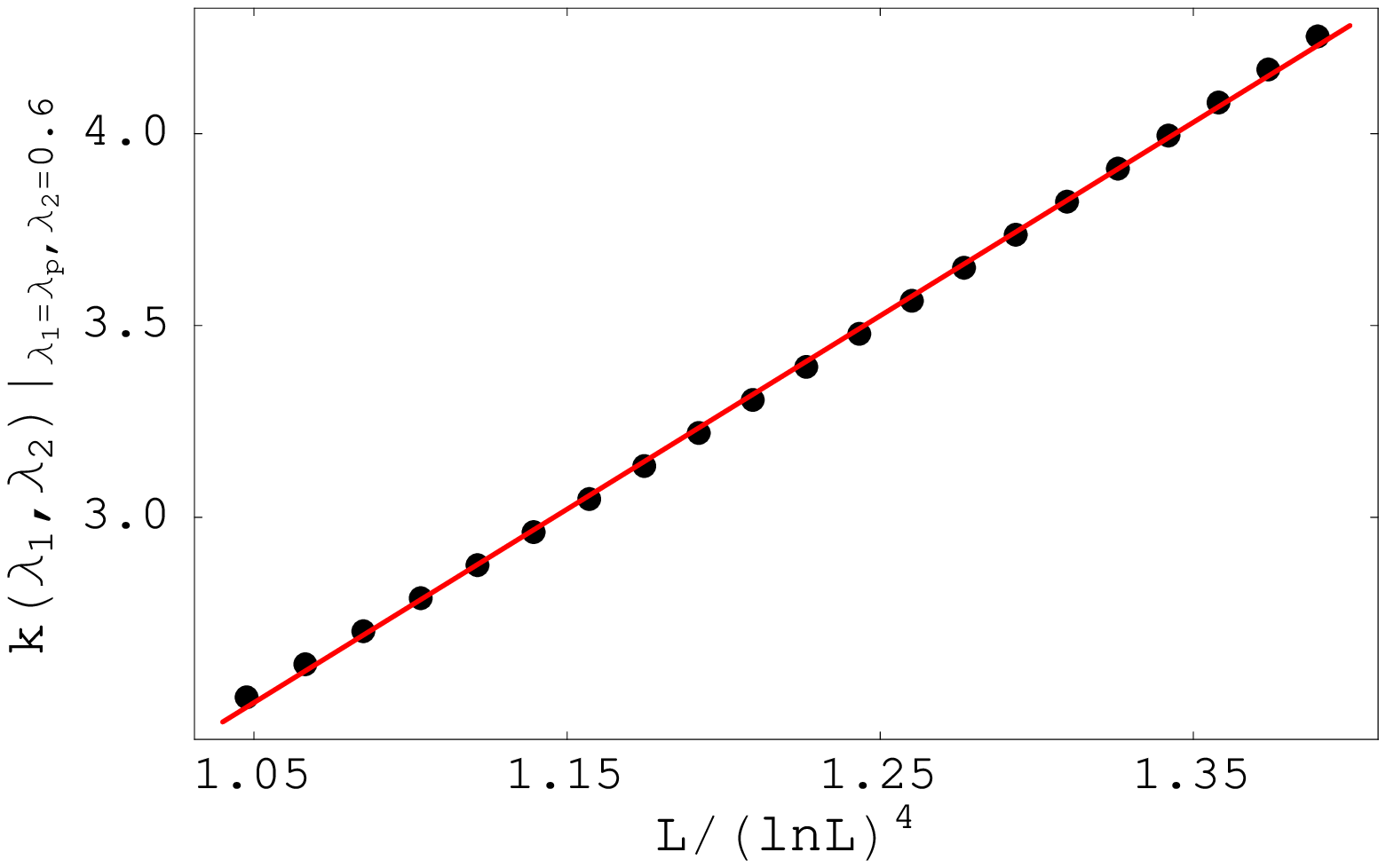}
\put(55,25){$(a)$}
\end{overpic}
\hspace{0in}
\begin{overpic}[width=40mm,totalheight=40mm]{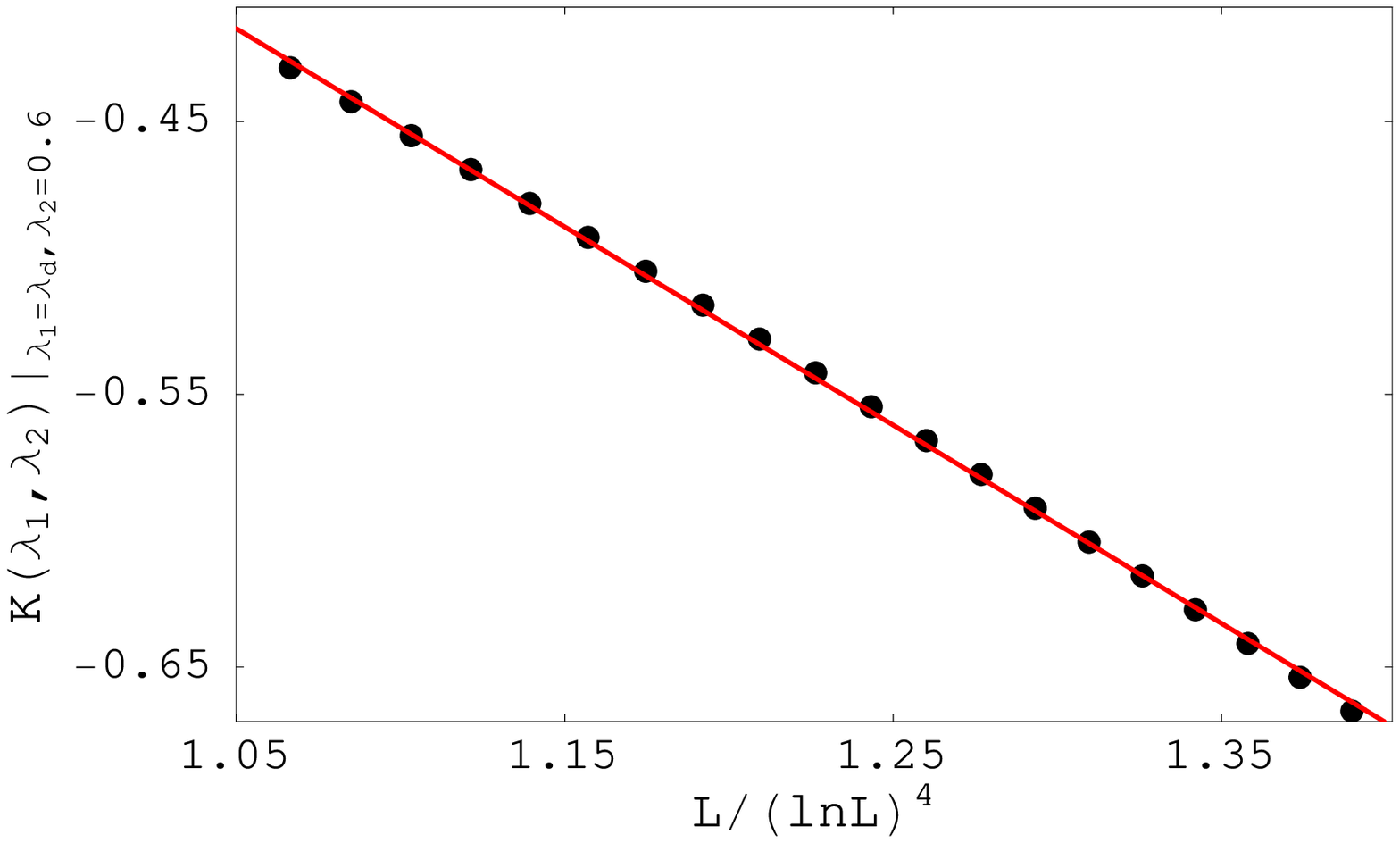}
\put(55,25){$(b)$}
\end{overpic}
\caption{(color online) (a) The peaks values of the Gaussian
curvature $K(\lambda_1,\lambda_2)$ of the quantum transverse Ising
model for large lattice sizes scale as $L/(\ln L)^4$. (b) The dips
values of the Gaussian curvature $K(\lambda_1,\lambda_2)$ of the
quantum transverse Ising model for large lattice sizes scale as
$L/(\ln L)^4$. In both cases, $\lambda_2 =0.6$ and $\gamma =1$.}
\label{fig2}
\end{figure}


{\it Finite size scaling analysis for the Gaussian curvature $K$.}
We focus on the quantum Ising universality class. The order
parameter, i.e., magnetization $\langle \sigma ^x \rangle$ is
non-zero for $\lambda < 1$, and otherwise zero. At the critical
point, the correlation length $\xi \sim |\lambda -\lambda_c
|^{-\nu}$ with $\nu = 1$~\cite{pfeuty}. In order to analyze how the
Gaussian curvature $K(\lambda_1,\lambda_2)$ behaves near the
critical point $\lambda_c =1$, we perform a finite size scaling
analysis for the quantum transverse Ising model.

\begin{figure}[ht]
\begin{overpic}[width=42mm,totalheight=42mm]{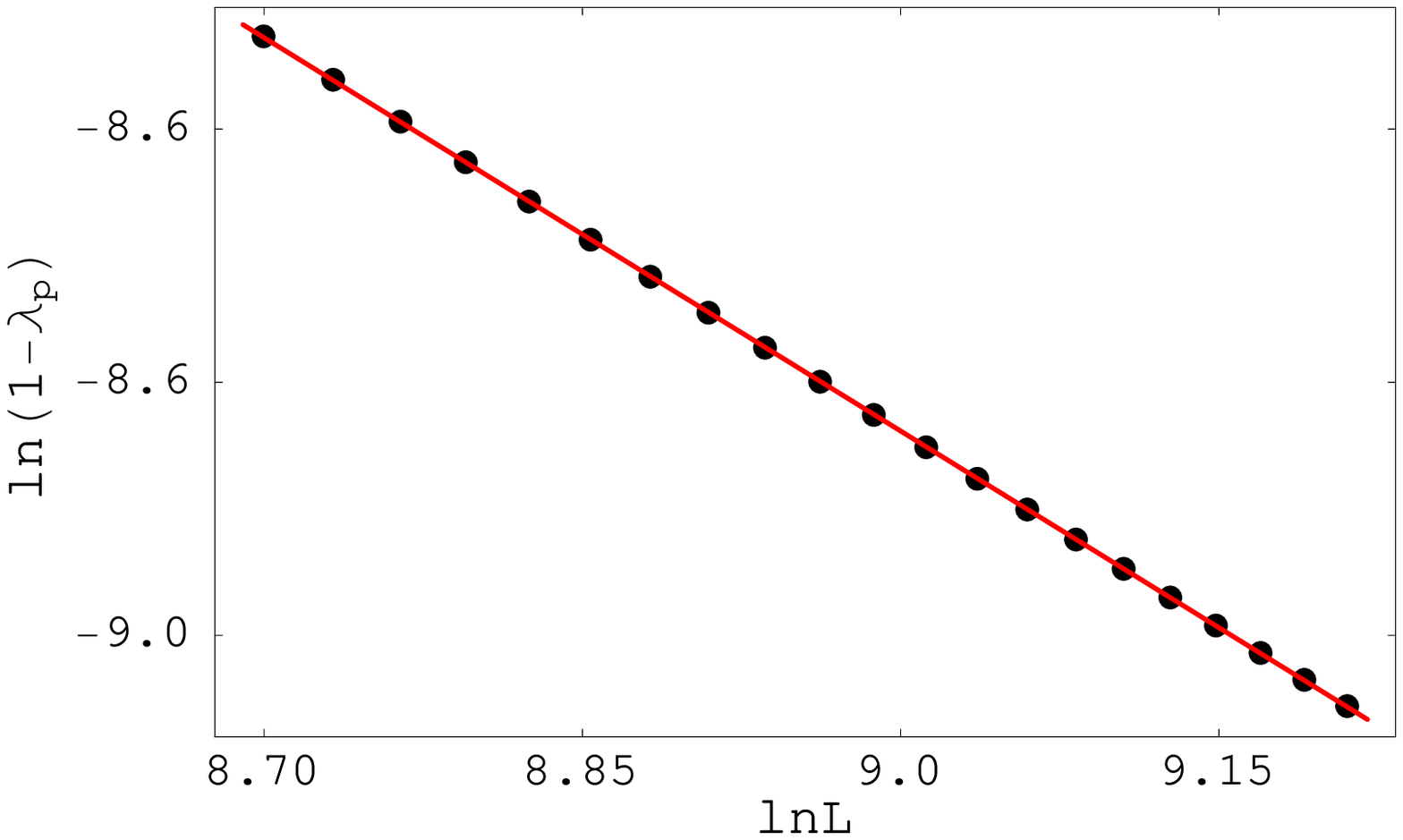}
\put(55,25){$(a)$}
\end{overpic}
\hspace{0in}
\begin{overpic}[width=40mm,totalheight=40mm]{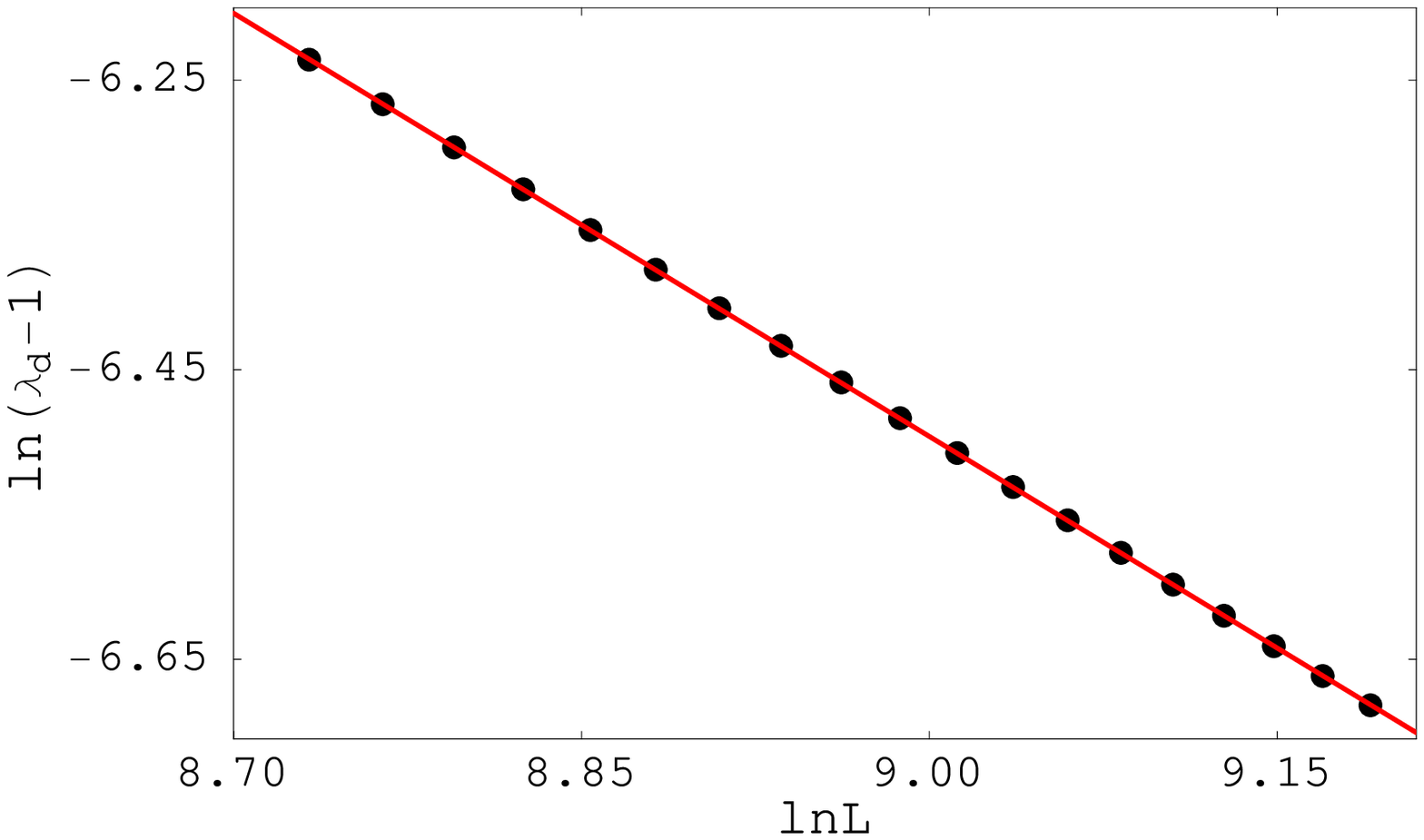}
\put(55,25){$(b)$}
\end{overpic}
\caption{(color online) (a): The positions of the peaks approach the
critical
  point $\lambda_c =1$ with increasing system size $L$ as $\lambda_p \approx 1-1.61490 L^{-1.03531}$.  (b) The positions of the dips approach the critical
  point $\lambda_c =1$  with increasing system size $L$ as $\lambda_d \approx 1+9.69198 L^{-0.974152}$. In both cases,  $\lambda_2 =0.6$ and $\gamma =1$.} \label{fig3}
\end{figure}

As already observed, the drastic change of the ground state wave
functions  makes the Gaussian curvature $K(\lambda_1,\lambda_2)$
divergent when the system undergoes the second order QPT at the
critical point $\lambda_c =1$ in the thermodynamic limit. However,
for finite-size systems, $K(\lambda_1,\lambda_2)$ remains to be
smooth for the quantum $XY$ model. In Fig.~{\ref{fig1}}, the
numerical results are also plotted for the Gaussian curvature
$K(\lambda_1,\lambda_2)$ with different system sizes, where
$\lambda_2=0.6$ and $\gamma=1$ (upper panel) and $\lambda_2=0.6$ and
$\gamma=1/2$ (lower panel). More precisely, in the thermodynamic
limit, $K(\lambda_1,\lambda_2)$ (as a function of $\lambda_1$ for a
fixed $\lambda_2$) diverges at the critical point
$\lambda_1=\lambda_c$:
\begin{equation}
K(\lambda_1,\lambda_2) \sim \frac{1}{|\lambda_1 -\lambda_c| (\ln
|\lambda_1 -\lambda_c|)^{4}} \label{infinite}.
\end{equation}
However, there is no divergence for finite-size systems, but there
are clear anomalies, featuring two quasi-critical values $\lambda_p$
and $\lambda_d$, one at each side of the critical point.   On the
left (right) side, the so-called quasi-critical points $\lambda_p$
($\lambda_d$) approach the critical value as $\lambda_p \approx
1-1.6149 L^{-1.03531}$($\lambda_d \approx 1+9.69198L^{-0.974152}$),
with the values at peaks (dips) diverging with increasing system
size $L$:
\begin{equation}
K(\lambda_1,\lambda_2)\big|_{\lambda_1=\lambda_{p(d)}} =
k_{p(d)}\frac{L}{(\ln L)^{4}} + {\rm constant}. \label{finite}
\end{equation}
Here the prefactor $k_{p(d)}$ is non-universal in the sense that it
depends on $\lambda_2$ and $\gamma$.  We emphasize that
Eq.~(\ref{finite}) follows from  Eq.~(\ref{infinite}), if we take
into account the fact that the model is conformally invariant at the
critical point. Indeed, on the one hand,  from Eq.~(\ref{infinite})
and the correlation length $\xi \sim |\lambda -\lambda_c |^{-\nu}$
with $\nu = 1$, we have $K(\lambda_1,\lambda_2) \sim \xi/(\ln
\xi)^4$.  On the other hand, the conformal invariance requires the
scale invariance: $\xi / L = \xi' / L'$.  The numerical results are,
respectively, plotted for
$K(\lambda_1,\lambda_2)|_{\lambda_1=\lambda_{p(d)}}$ in
Fig.~\ref{fig2} and  for $\lambda_{p(d)}$ in Fig.~\ref{fig3} with
$\lambda_2 = 0.6$ and $\gamma = 1$.  The same is also true for any
nonzero $\gamma$.  This shows that, consistent with the exact
result, the correlation length critical exponent $\nu$ equals 1, as
long as $\gamma$ is nonzero.

{\it Conclusions.} We have shown that singularities in fidelity
surfaces may be \textit{intrinsically} characterized in terms of
Riemannian geometry, based on the fidelity description of QPTs.
Generically, the Ricci curvature tensor for finite-size systems is
analytic and it exhibits singularities at transition points in the
thermodynamic limit, as reflected in the Ricci scalar curvature that
blows up when the system size tends to $\infty$. This opens up the
possibility to exploit the theory of Ricci flows~\cite{perelman} to
characterize QPTs in condensed matter theory. The one-dimensional
quantum Ising model in a transverse field is exploited as an example
to explicitly illustrate the theory~\cite{explanation}, and a finite
size scaling analysis has been performed for the Ricci scalar
curvature with different lattice sizes, and the correlation length
critical exponent has been extracted, consistent with the known
exact value.

We thank John Paul Barjaktarevi\v{c}, Sam Young Cho and John
Fjaerestad for helpful discussions and comments. The support from
the National Natural Science Foundation of China is acknowledged.


\begin{thebibliography}{99}
\bibitem{finitefidelity} The explicit
expression for $\ln d(\lambda_1,\lambda_2)$ was also given in
Refs.~\cite{zjp,zhou} for the quantum $XY$ model of finite sizes,
which is needed for the evaluation of the Gaussian curvature
$K(\lambda_1,\lambda_2)$ for the corresponding fidelity surfaces.
\end{thebibliography}
\end{document}